\documentclass[universe,article,accept,pdftex,moreauthors]{Definitions/mdpi} 

\firstpage{1} 
\pubvolume{1}
\issuenum{1}
\articlenumber{1}
\pubyear{2026}
\copyrightyear{2026}
\externaleditor{Firstname Lastname}
\datereceived{9 January 2026} 
\daterevised{2 February 2026} 
\dateaccepted{6 February 2026} 
\datepublished{ } 

\usepackage{dcolumn}
\usepackage{bm}
\usepackage{arydshln}
\usepackage[english]{babel}
\usepackage{longtable}
\usepackage{float}
\usepackage[utf8]{inputenc}
\usepackage{array,multirow}
\usepackage{color}
\usepackage{appendix}
\usepackage{makeidx}
\usepackage{graphicx}
\usepackage{amsmath}
\usepackage{bbm}
\usepackage{yfonts}
\usepackage{amsfonts}
\usepackage{amssymb}
\usepackage{fancyhdr}
\usepackage{graphicx}
\usepackage{setspace}
\usepackage{soul}
\usepackage{physics}
\usepackage{makecell,tabularx}
\theoremstyle{remark}
\newcommand{\be}{\begin{equation}}
\newcommand{\ee}{\end{equation}}
\def\dif{{\rm d}}

\Title{Thermodynamic Interpretation of the Kompanneets--Chernov--Kantowski--Sachs Solutions} 

\Author{Salvador Mengual
 $^{1}$ and Joan Josep Ferrando $^{1,2,}$*}

\AuthorNames{Salvador Mengual and Joan Josep Ferrando}

\address{$^{1}$ \quad Departament d'Atronomia i Astrofísica, Universitat de València, 46100 Burjassot, València, Spain; salvador.mengual@uv.es 
\\
$^{2}$ \quad Observatori Astronòmic, Universitat de València, 46980 Paterna, València, Spain}

\corres{Correspondence: joan.ferrando@uv.es}

\abstract{The spatially homogeneous perfect fluid solutions by Kompanneets--Chernov--Kantowski--Sachs are interpreted as a thermodynamic perfect fluid in isentropic evolution, namely, the isentropic limit of their non-homogeneous generalizations, the T-models. Some specific solutions that model a generic ideal gas are examined, and the associated thermodynamic variables are obtained. We show that the necessary macroscopic conditions for physical reality are fulfilled in wide spacetime domains. The field equations for a classical ideal gas are established, and the behavior of the solution is analyzed. The models fulfilling a relativistic $\gamma$-law are also examined, and the solutions for some particular cases are obtained.}

\keyword{thermodynamic solutions; homogeneous anisotropic models} 

\begin{document}


\section{Introduction} \label{sec-intro}

Kompaneets--Chernov--Kantowski--Sachs (KCKS) solutions are homogeneous but aniso\-tropic cosmological solutions to the Einstein equations. Some physical properties of these solutions were investigated by Kompaneets and Chernov~\cite{Kompaneets-Chernov} and they were studied with more detail by Kantowski and Sachs for dust~\cite{Kantowski-Sachs}. The subfamily of spherically symmetric solutions are spatially homogeneous metrics that do not permit a simply transitive group of motions, and the subspaces of constant time coordinate do not contain their centers of symmetry. However, the counterparts with plane and hyperbolic symmetry are of Bianchi types I and III, respectively, so they do have simply transitive three-dimensional subgroups~\cite{Inhomogeneous_Cosmological_Models, Krasinski-Plebanski, Ellis-Maartens-MacCallum, Kramer}.

Several works have been devoted to studying the behavior of the KCKS metric, motivated by the interest in spatially homogeneous cosmological models~\cite{Weber-1984, Weber-1986} (see also~\cite{Inhomogeneous_Cosmological_Models} and references therein). Interest in this family of solutions remains strong due to the observations of the cosmic microwave background isotropy and of the spatial distribution of galaxies at large scales. The stability and evolution of the anisotropies of the KCKS metric have been studied~\cite{Li-Hao-2003} and they have been used to model inflationary scenarios in which the initial anisotropies die away~\cite{Mendes-Enriques-1991, Ghorani-Heydarzade-2021}. Moreover, these solutions have also been used to model gravitational collapse~\cite{Terezon-Campos-2021}. 

Through all these works, different fluid sources of the field equations have been considered: dust fluid, perfect fluid, anisotropic fluid, fluid with bulk viscosity, and Skyrme fluid (see~\cite{Tiwari-Dwivedi-2008, Chakraborty-Roy-2008, Adhav-2011, Keresztes-2015, Parisi-2015}). However, most of the attention has been focused on the spacetime behavior, rather than on the source behavior, and its interpretation as a physically realistic fluid. Many of the solutions already studied have no clear physical meaning~\cite{Vajk, McVittie, Herlt} and some authors have pointed out the difficulties in associating a realistic equation of state to these solutions and their non-homogeneous generalizations, the T-models and the Szekeres--Szafron type II solutions~\cite{Lima-Tiomno, Bolejko}. Nevertheless, some recent works have provided interesting thermodynamic interpretations of these families~\cite{Krasinski-et-al, C-F-S_SzSz_Singular, C-F-S_SzSz_Regular, FM-Termo-T-models, FM-General-sol-T-models}.


In this paper, we present a further step in this direction by giving a thermodynamic interpretation of the KCKS solutions. We show that every KCKS model can be interpreted as the isentropic evolution of a thermodynamic fluid whose non-isentropic evolution is represented by a T-model. Moreover, we also study the KCKS solution that can represent the isentropic evolution of a classical ideal gas, as well as those that fulfill a relativistic $\gamma$-law barotropic relation.

In the following subsection we explain the framework within which this study takes place and the methodology used in it.

\subsection{Macroscopic Necessary Conditions for Physical Reality}
\label{subsec-necessary-conditions}

Most of the perfect fluid solutions to Einstein equations have been derived without specifying an equation of state, while others have been obtained in the dust case or by prescribing a (non-physical) time-dependence of the pressure, or also by imposing particular barotropic relations to close the system of equations. While these approaches have led to a broad spectrum of exact solutions~\cite{Kramer, Inhomogeneous_Cosmological_Models, Griffiths-Podolsky-2009}, many of them lack a clear physical interpretation. Consequently, studying the potential physical viability of these metrics is an important task in the research on exact solutions.

Inspired in a paper by Coll and Ferrando~\cite{Coll-Ferrando-termo}, our research group has developed an approach to the perfect fluid hydrodynamics~\cite{Hydro-LTE, Coll_Ferrando_i_Saez_2020b}, which has proved to be a useful tool in analyzing the physical meaning of specific families of perfect fluid solutions to Einstein's equations~\cite{CF-Stephani, C-F-S_SzSz_Singular, C-F-S_SzSz_Regular, CFS-CIG, FM-Termo-T-models, FM-General-sol-T-models, FM-R-models-plans, SFM-Stephani}. Next, we summarize the main ideas of this approach.

The evolution of a relativistic perfect fluid is described by an energy tensor in the form $T = (\rho+ p) u \otimes u + p \, g$, and submitted to the conservative condition $\nabla \cdot T=0$. This constraint consists of a differential system of four equations on five {\em hydrodynamic quantities} (three components of the {\em unit velocity} $u$, {\em energy density} $\rho$, and {\em pressure} $p$):
\be \label{ceq}
\hspace{-5mm} {\rm C} :  \qquad \qquad  \dif p  + u(p) u + (\rho + p) a = 0 \, ,  \qquad 
u(\rho) + (\rho+ p) \theta = 0 \, ,
\ee
where $a$ and $\theta$ are, respectively, the acceleration and the expansion of $u$, and where $u(q)$ denotes the directional
derivative, with respect to $u$, of a quantity $q$, $u(q)
= u^{\alpha} \partial_{\alpha} q$. 

However, for $T$ to represent the energetic evolution of a physically realistic fluid, complementary general macroscopic requirements must be imposed on it. 

Pleba\'nski~\cite{Plebanski_1964} {\em energy conditions} are necessary algebraic conditions for physical reality and, in the perfect fluid case, they state the following: 
\begin{equation} \label{e-c}
\hspace{-5mm} {\rm E} : \qquad \qquad  \qquad \qquad  -\rho < p \leq \rho  \, . \qquad \qquad \qquad 
\end{equation}

Furthermore, if we want to describe the (non-isoenergetic, $u(\rho) \not= 0$) evolution of a thermodynamic perfect fluid in {\em local thermal equilibrium}, the hydrodynamic quantities $\{u, \rho, p\}$ must fulfill the {\em hydrodynamic sonic condition}~\cite{Coll-Ferrando-termo, Hydro-LTE}: 
\begin{equation} \label{lte-chi}
\hspace{-5mm} {\rm S} :  \qquad \qquad     \dif \chi \wedge \dif p \wedge \dif \rho = 0 \, , \qquad \chi \equiv \frac{u(p)}{u(\rho)}   \, .
\end{equation}
When $\dif \rho \wedge \dif p \not=0$, the above condition is tantamount to the usual thermodynamic deterministic closure for the differential system C. Indeed, when S holds, the {\em indicatrix of the local thermal equilibrium} $\chi$ is a function of state, $\chi = \chi(\rho,p)$, which physically represents the square of the {\em speed of sound} in the fluid, $\chi (\rho ,p) \equiv  c^2_{s}$. Moreover, a set $\{n, \epsilon, s, \Theta\}$ of {\em thermodynamic quantities} ({\em matter density} $n$, {\em specific internal energy} $\epsilon$, {\em temperature} $\Theta$, and {\em specific entropy} $s$) exists, which is constrained by the common thermodynamic laws~\cite{Eckart, Rezzolla}. 
Namely, the conservation of matter:
\begin{equation}  
\nabla \cdot (nu) = u(n) + n \theta = 0 \, ,  \label{c-masa}
\end{equation}
the {\em local thermal equilibrium relation}, which can be written as
\begin{equation}
\Theta \dif s = \dif h - \frac{1}{n} \dif p \, ,  \qquad h \equiv \frac{\rho+p}{n} \, , \label{re-termo}
\end{equation}
where $h$ is the {\em relativistic specific enthalpy}, and the decomposition defining the specific internal energy:
\begin{equation}
\rho= n(1+\epsilon) \, .  \label{masa-energia} 
\end{equation}

When the conservation equations C and the hydrodynamic sonic condition S hold, we say that $T\equiv \{u, \rho, p\}$ defines the {\em hydrodynamic flow} of a thermodynamic perfect fluid in local thermal equilibrium. Then, the family of {\em thermodynamic schemes} $\{n, \epsilon, s, \Theta\}$ associated with a hydrodynamic flow $T\equiv \{u, \rho, p\}$ is obtained as follows (inverse problem)~\cite{Hydro-LTE}: the specific entropy $s$ and the matter density $n$ are of the form $s= s(\bar{s})$ and $n= \bar{n}R(\bar{s})$, where $s(\bar{s})$ and $R(\bar{s})$ are arbitrary real functions of a particular solution $\bar{s}=\bar{s}(\rho, p)$ to the equation $u(s)=0$, and $\bar{n}=\bar{n}(\rho,p)$ is a particular solution to Equation (\ref{c-masa}). Moreover, $\Theta$ and $\epsilon$ are determined, respectively, by (\ref{re-termo}) and (\ref{masa-energia}).

On the other hand, the matter density, the temperature, and the specific internal energy must be positive:
\begin{equation} \label{P}
\hspace{-5mm} {\rm P} :  \qquad \qquad \qquad    \Theta > 0 \, , \qquad  \qquad   \rho > n > 0   \, . \qquad \qquad
\end{equation}

Finally, in order to obtain a coherent theory of shock waves for the {\em fundamental system of perfect fluid hydrodynamics} \{(\ref{ceq}), (\ref{c-masa}), (\ref{re-termo}), (\ref{masa-energia})\}, one must require the relativistic compressibility conditions ~\cite{Israel_1960, Lichnerowicz_1966}. They impose the inequalities ${\rm H}_1$: $(\tau'_p)_s < 0, \ (\tau''_p)_s > 0$, and the inequality ${\rm H}_2$: $(\tau'_s)_p > 0$, where the function of state $\tau = \tau(p, s)$ is the {\em dynamic volume}, with $\tau = \hat{h}/n$, $\hat{h} = h/c^2$ being the dimensionless enthalpy index. In~\cite{Coll_Ferrando_i_Saez_2020b}, we showed that the compressibility conditions H$_1$ only restrict the hydrodynamic quantities, and that they can be stated in terms of the function of state $c_s^2 = \chi(\rho,p)$: 
\begin{equation}
\hspace{-5mm} {\rm H}_1 : \qquad \qquad 0 < \chi < 1 \, , \qquad    (\rho+p)(\chi \chi_{p}' + \chi_{\rho}') + 2 \chi(1-\chi) > 0   \, .       \label{cc-1-chi}
\end{equation}
However, compressibility condition H$_2$ imposes constraints on the thermodynamic scheme and it can be stated as follows~\cite{Coll_Ferrando_i_Saez_2020b}:
\be \label{H2-Theta}
\hspace{-5mm} {\rm H}_2 : \qquad \qquad \qquad  2 n \Theta > \frac{1}{s_{\rho}'} \, .  \qquad \qquad \qquad 
\ee
In the above two expressions, and in what follows, we use the notation $f_{\rho}' \equiv (\frac{\partial f}{\partial \rho})_p$ and $f_{p}' \equiv (\frac{\partial f}{\partial p})_\rho$ for any function $f = f(\rho, p)$. 

Note that in the general necessary macroscopic constraints C, S, E, P, H$_1$, and H$_2$ specified above, we must distinguish two types of conditions according to their nature:

\begin{enumerate}[label=,labelsep=7.5mm]
\item[(i)] {\em Hydrodynamic constraints}: the conservation equation C, the energy conditions E, the hydrodynamic sonic condition S, and the compressibility conditions H$_1$ exclusively involve the hydrodynamic quantities $\{u, \rho, p\}$. They fully determine
the hydrodynamic flow of the thermodynamic fluid in local thermal equilibrium, whether the fluid is treated as a test fluid or as the source of the gravitational field. In the latter case, they also constrain the admissible gravitational field through the Einstein equations when looking for perfect fluid solutions.

\item[(ii)] {\em Thermodynamic constraints}: the positivity conditions P and the compressibility condition H$_2$ restrict the thermodynamic schemes $\{n, \epsilon, s, \Theta\}$ associated with a
hydrodynamic flow $\{u, \rho, p\}$. These schemes offer different physical interpretations for a given hydrodynamic perfect fluid flow. Moreover, when looking for solutions to Einstein’s equations, these conditions do not restrict the admissible gravitational field.
\end{enumerate}


\subsection{Barotropic Perfect Energy Tensor}

In the above subsection, we analyzed the sonic condition (\ref{lte-chi}) for a non-barotropic perfect energy tensor, $\dif \rho \wedge \dif p \not=0$. This generic case can be useful to understand the thermodynamic approach to the T-models presented in~\cite{FM-Termo-T-models, FM-General-sol-T-models} and used in the present paper. 

However, the KCKS solutions that we study here have a barotropic perfect energy tensor, that is,  $\dif \rho \wedge \dif p =0$. Then, the sonic condition S also holds, and we must clarify what this condition physically means. In fact, we can consider two possible interpretations. 

When we have a non-constant energy density, $d \rho \not= 0$,
the barotropic condition can be stated as a barotropic relation of the form
\be \label{p-rho}
p = \phi(\rho).
\ee
This barotropic relation can be interpreted as an equation of state of the medium, which holds independently of the considered particular evolution $T$. Then, we say that this medium is an (intrinsically) barotropic perfect fluid. But it can also be interpreted as a particular evolution of non (intrinsically) barotropic media. For example, if a fluid with an equation of state $p = p(\rho, s)$ evolves at constant entropy $s_0$, we have $p = p(\rho, s_0) \equiv \phi(\rho)$ for this particular evolution.

Solving the inverse problem for a barotropic perfect energy tensor means determining both the associated (intrinsically) barotropic thermodynamic schemes, as well as the non-barotropic fluids that can follow this specific barotropic evolution.

Concerning the intrinsically barotropic media, for which (\ref{p-rho}) is an equation of state, the associated thermodynamic schemes and physically significant examples can be found in~\cite{Hydro-LTE}. For instance, $\rho=3p$ models an ultrarelativistic fluid.

%
%
%

When considering (\ref{p-rho}) as an evolution equation, the richness of possible physical interpretations is very broad~\cite{Hydro-LTE}. Now, we will restrict ourselves to the generic ideal gases in isentropic evolution. These media meet the equation of state $p = \tilde{k} n \Theta$, where $\tilde{k} \equiv k_B / m$, with $k_B$ being the constant's Boltzmann and $m$ the mass of the particles. Furthermore, they can be characterized by the {\em ideal sonic condition}~\cite{Hydro-LTE}:
\begin{equation}
\chi = \chi(\pi) \not= \pi, \qquad \chi \equiv \frac{u(p)}{u(\rho)} , \qquad \pi \equiv \frac{p}{\rho}  \,  .  \label{gas-ideal-sonic}
\end{equation}
Moreover, the associated ideal thermodynamic scheme has a specific entropy $s$, a matter density $n$, and a temperature $\Theta$, given by~\cite{Hydro-LTE}
\begin{eqnarray}
s(\rho,p) = \tilde{k} \ln \frac{f(\pi)}{\rho} \, , \qquad n(\rho,p) = \frac{\rho}{e(\pi)} \, , \qquad  \Theta(\rho, p) = \frac{\pi}{\tilde{k}}\, e(\pi) \, ;
\label{ideal-n-T} \\
f(\pi) \equiv f_0 \exp \left\lbrace \! \int \! \! \frac{\dif \pi}{\chi(\pi)-\pi}\right\rbrace   \,  , \qquad e(\pi) \equiv e_0 \exp \left\lbrace \! \int \! \! \frac{\pi \dif \pi}{(\pi+1)(\chi(\pi)-\pi)}\right\rbrace   \,  .
\label{ideal-s}
\end{eqnarray}
Consequently, an isentropic evolution, $s(\rho,p) = s_0$, leads to an implicit barotropic relation of the form 
\begin{equation} \label{rho-pi}
\rho = K f(p/\rho) \, , \qquad K \equiv \exp{-s_0/\tilde{k}} \, .
\end{equation}

In the case of a classical ideal gas, $\epsilon = C_v \Theta$, the barotropic relation (\ref{rho-pi}) becomes~\cite{CFS-CIG, FS-KCIG}
\begin{equation} \label{CIG-isentropic}
	(\gamma - 1) \rho = p + B p^{1/\gamma}, \qquad B \equiv exp \left\lbrace \frac{s_0(\gamma - 1)}{\tilde{k} \, \gamma} \right\rbrace \, ,
\end{equation}
where $\gamma$ is the adiabatic index. Moreover, the matter density $n$ is given by
\be \label{n-CIG}
n = \frac{B}{\gamma-1}p^{1/\gamma}.
\ee

And for the Taub--Mathews approximation to the Synge gas~\cite{Rezzolla, FM-Synge}, we obtain the implicit barotropic relation
\begin{equation} \label{isentropic-TM}
	\tilde{B} p^3 = \rho (\rho - 3p)^4 , \qquad \tilde{B} \equiv \exp \left\lbrace \frac{2 s_0}{\tilde{k}} \right\rbrace \, .
\end{equation}

\subsection{About This Paper}

In Section \ref{sec-KCKS-T-models}, we develop the main part of the paper by showing that every KCKS metric represents the isentropic evolution of a thermodynamic T-model. First, we introduce the notation and we summarize some results about the thermodynamic approach to the T-models given in~\cite{FM-Termo-T-models}. Here, we present these already known results in a more suitable way to this work. It is worth remarking that the solutions depend on an arbitrary function $Q(r)$ of the radial coordinate and an arbitrary function $\varphi(t)$ of the time coordinate. In the expression of the metric line element we can also find a function $v(t)$, which can be fixed by an adequate choice of the time coordinate, and two functions $\alpha_i(t)$, which are determined by the field equations. Second, we study the KCKS limit, which is reached when $Q(r) = constant$. This condition implies a constant entropy and, consequently, every KCKS solution represents the isentropic evolution of a family of T-models. Third, we consider in detail the solutions that model a generic ideal gas and we analyze the behavior of the thermodynamic quantities.  

Section \ref{sec-KCKS-other} is devoted to analyzing whether the KCKS solutions can model other fluids different from those represented by T-models. On the one hand, we state the field equations when the barotropic relation of a CIG in isentropic evolution is imposed. The case of plane symmetry is solved numerically, and the physical behavior of these Bianchi type I models is studied and compared with the FLRW limit. On the other hand, we establish the field equations for the relativistic $\gamma$-law models and we solve them for the case of plane symmetry by using an integration algorithm given in~\cite{FM-General-sol-T-models}.

Finally, in Section \ref{sec-conclusions}, we comment about our results and possible further work.


\section{The KCKS Solutions as the Isentropic Evolution of a T-Model} \label{sec-KCKS-T-models}
	
	The T-models are the perfect fluid solutions to the Einstein equations admitting a three-dimensional group G$_3$ of isometries acting on space-like two-dimensional orbits S$_2$ whose curvature gradient is tangent to the fluid flow~\cite{Ruban_1969, Krasinski-Plebanski}. 	
	
The spherical dust T-model was first considered by Datt~\cite{Datt_1938}, and the dust solution with cosmological constant was widely analyzed later by Ruban, who showed that this solution has no Newtonian analogue~\cite{Ruban_1968, Ruban_1969}. The perfect fluid T-models with a non-constant pressure were examined by Korkina and Martinenko~\cite{Korkina-Martinenko_1975} and Ruban~\cite{Ruban_1983}, while Herlt~\cite{Herlt} proposed an algorithm to obtain new solutions in this
family (see also~\cite{Kramer}). 	

The KCKS metrics correspond to the spatially homogeneous limit of the T-models.	 This fact enables us to interpret the KCKS solutions as the isentropic evolution of a T-model.
	
\subsection{Thermodynamics of the T-Models}

It is known~\cite{Ruban_1969} that the spherical perfect fluid T-models have geodesic motion, a result that can be extended to the plane and hyperbolic symmetries (see, for example,~\cite{Krasinski-Plebanski}). In co-moving--synchronous coordinates, its line element can be written as~\cite{Kramer}
\begin{eqnarray} \label{metric-ss-1}
	ds^2 = -e^{2\nu(t)} dt^2 + e^{2\lambda(t,r)} dr^2 + Y^2(t) C^2 (dx^2 + 	dy^2),  	\ \\[3mm]
	C = C(x,y) \equiv \left[1 + \frac{k}{4}(x^2 + y^2)\right]^{-1} \!\!\!\!, \quad k = 0, \pm 1 , \ \label{metric-ss-2}
\end{eqnarray}
where the value of $k$ distinguishes the plane, spherical, and hyperbolic symmetries. In~\cite{FM-Termo-T-models}, we study when these solutions represent the evolution in local thermal equilibrium of a fluid that meets the suitable macroscopic physical constraints displayed in the introduction. In~\cite{FM-General-sol-T-models}, we show that, if we make $e^{2 \lambda} = \varphi(t) \alpha^2(t,r) > 0$, $e^{-2\nu} = v(t) > 0$ and $Y^2 = \varphi (t) > 0$, the field equation that these metric functions need to verify so that (\ref{metric-ss-1}) is a perfect fluid solution is
\begin{equation} \label{eq-T-1}
2 v \varphi \, \ddot{\alpha} + (\dot{v} \varphi + 3v \dot{\varphi})\, \dot{\alpha} -  2k \, \alpha  = 0 \, ,
\end{equation}
where a dot denotes derivative with respect to the time coordinate $t$. Moreover, the unit velocity of the fluid $u = \sqrt{v} \, \partial_t$ is geodesic and its expansion is
\begin{equation} \label{expansion-T-1}
\theta = \sqrt{v} \left(\frac32 \frac{\dot{\varphi}}{\varphi} + \frac{ \dot{\alpha}}{\alpha}\right)  = \sqrt{v} \, \partial_t [\ln(\varphi^{3/2} \alpha)] \,  .
\end{equation}
And the {\em pressure} $p$ and the {\em energy density} $\rho$ are then given by
\begin{eqnarray} \label{pressure-T-1}
p =  v \left[\frac14 \frac{\dot{\varphi}^2}{\varphi^2} - \frac{\ddot{\varphi}}{\varphi} - \frac12 \frac{\dot{\varphi}}{\varphi}  \frac{ \dot{v}}{v}\right] - \frac{k}{\varphi} \, , \\[1mm]
\rho =  v \left[\frac34 \frac{\dot{\varphi}^2}{\varphi^2} +  \frac{\dot{\varphi}}{\varphi} \frac{\dot{\alpha}}{\alpha}\right] + \frac{k}{\varphi} \, .
\label{density-T-1}
\end{eqnarray}

Notice that (\ref{eq-T-1}) is a second-order linear differential equation for the function $\alpha(t,r)$ when $v(t)$ and $\varphi(t)$ are given. Consequently, we can change the coordinate $r$ so that its general solution is of the following form:
\be \label{w-w1-w2}
\alpha(t,r) =  \alpha_1(t) + \alpha_2(t) \, Q(r) \, ,
\ee
where $Q(r)$ is an arbitrary real function, with $\alpha_i(t)$ being two particular solutions to Equation~(\ref{eq-T-1}).

As we have the freedom to choose the coordinate $t$ without changing the spacetime metric, we take $v(t)=1$, and then the coordinate $t= \tau$ is the proper time of the co-moving observer. We could have imposed any other condition on the time-dependent metric functions to fix this election. We do this in~\cite{FM-General-sol-T-models} conveniently in order to find the general solution to (\ref{eq-T-1}). 

With this election of the time coordinate, we studied, in~\cite{FM-Termo-T-models}, the thermodynamics of the T-models. In that paper, we worked with the coordinate function $\omega = \alpha \sqrt{\varphi}$ instead of $\alpha$. Now, we summarize the results in terms of $\alpha$.  All of the T-models fulfill the sonic condition (\ref{lte-chi}), and the indicatrix function is of the following form:
\be  \label{chi-Tmodels}
c_s^2  =  \chi(\rho,p)  \equiv \frac{1}{{\cal A}(p) \rho^2 + {\cal B}(p) \rho + {\cal C}(p)} \, ,
\ee
where ${\cal A}$, ${\cal B}$, and ${\cal C}$ are functions of $\tau$ (and then of $p$) through the metric functions $\varphi, \alpha_i$ and their derivatives. Moreover, the metric function $Q(r)$ is a function of state given by
\begin{equation} \label{Q(prho)}
Q = - \frac{\psi(\tau) \, \alpha_1 + \varphi \, \dot{\varphi} \, \dot{\alpha}_1}{\psi(\tau) \, \alpha_2 + \varphi \, \dot{\varphi} \, \dot{\alpha}_2} \equiv Q(\rho,p)\, \qquad \psi(\tau) \equiv \frac34 \dot{\varphi}^2 + k \, \varphi - \rho \, \varphi^2  \, .
\end{equation}

On the other hand, the thermodynamic schemes associated with the T-models are determined by a specific entropy $s$ and a matter density $n$ of the following form:
\begin{equation}  \label{s-n-Tmodels}
s(\rho, p) = s(Q)\, , \quad  n(\rho,p) = \frac{1}{\varphi^{3/2}(\alpha_1 + \alpha_2 Q)N(Q)}\, ,
\end{equation}
where $s(Q)$ and $N(Q)$ are two arbitrary real functions of the function of state $Q(r)$. Moreover, the temperature of the thermodynamic scheme defined by each pair $\{s, n\}$ takes \mbox{the expression}
\begin{equation} \label{T-Tmodels}
\Theta = \ell(Q) \lambda_1(\tau) + m(Q) \lambda_2(\tau) \equiv \Theta(\rho,p) \, ,
\end{equation}
\begin{equation} \label{lambda_i}
\lambda_i(\tau) \equiv \sqrt{\varphi}\left[\dot{\varphi}\, \dot{\alpha}_i + (\frac{\dot{\varphi}^2}{\varphi}- \ddot{\varphi}) \alpha_i\right]  \, ;
\end{equation}
\begin{equation} \label{ell-m}
\ell(Q) \equiv \frac{N'(Q)}{s'(Q)} , \quad  m(Q) \equiv \frac{1}{s'(Q)}[Q N'(Q) + N(Q)] .
\end{equation}

Understanding the physical meaning of these T-models implies analysis of the spacetime domains where a specific solution satisfies the energy conditions E and the compressibility conditions H$_1$, and the associated thermodynamic schemes fulfill the positivity conditions P and the compressibility condition H$_2$ (see~\cite{FM-Termo-T-models} for more details).

\subsection{Isentropic Limit of a T-Model} 

 The KCKS metrics are the spatially homogeneous limit of the T-models. They can also be characterized by one of the following three equivalent conditions: (i) the metric function $\alpha(t,r)$ factorizes, and then one can take the coordinate $r$ so that $\alpha= \alpha(t)$ (that is, $Q(r)=Q_0 = constant$); (ii) the energy density is homogeneous, $\rho = \rho(t)$; and (iii)  the fluid expansion is homogeneous, $\theta = \theta(t)$.

Each election of the arbitrary metric function $Q(r)$ defines a different particular T-model. Each of these particular solutions, in turn, has an associated set of thermodynamic schemes, given by (\ref{Q(prho)})--(\ref{ell-m}). The different thermodynamic schemes correspond to the different elections of the arbitrary functions $s(Q)$ and $N(Q)$. 

Now, the isentropic limit of these thermodynamic schemes is achieved by making $Q = Q_0 =constant$, which, therefore, corresponds to the spatially homogeneous KCKS limit. Consequently, the KCKS solutions can be interpreted as the isentropic limit of the T-models. 

With the election of the proper time $\tau$ as the time coordinate ($v(\tau)=1$), the metric line element of a KCKS solutions takes the following expression:
\begin{equation} \label{metric-KCKS}
ds^2= -d \tau^2 + \varphi(\tau)\{[\alpha_1(\tau) + Q_0\, \alpha_2(\tau)]^2 dr^2 +  C^2 (dx^2+dy^2)\},
\end{equation}
where $C$ is given in (\ref{metric-ss-2}), $\varphi(\tau)$ is an arbitrary function, $Q_0$ is a real parameter, and $\alpha_i(\tau)$ are two particular solutions of the field equation:
\begin{equation} \label{eq-KCKS}
2 \varphi \, \ddot{\alpha} + 3 \dot{\varphi}\, \dot{\alpha} -  2k  \,\alpha = 0 \, .
\end{equation}
Note that the space of KCKS solutions is controlled by $\{Q_0, \varphi(\tau)\}$, a real parameter and a real function of the time coordinate.

Now, the {\em pressure} $p$ and the {\em energy density} $\rho$ become
\begin{equation} \label{pressure-density-KCKS}
p =  \frac14 \frac{\dot{\varphi}^2}{\varphi^2} - \frac{\ddot{\varphi}}{\varphi} - \frac{k}{\varphi} \, , \qquad \quad
\rho = \frac34 \frac{\dot{\varphi}^2}{\varphi^2} +   \frac{\dot{\varphi}(\dot{\alpha}_1 + Q_0\, \dot{\alpha}_2)}{\varphi(\alpha_1 + Q_0\, \alpha_2)} + \frac{k}{\varphi} \, ,
\end{equation}
and the constant $Q_0$ is the function of state given in (\ref{Q(prho)}).  
Note that $Q(\rho,p)$ depends explicitly on $\rho$ and implicitly on $p$ though the metric functions $\varphi, \alpha_i$. Condition \mbox{$Q_0 = Q(\rho,p)$} gives us an (implicit) barotropic relation, which follows from the imposed isentropic evolution. When considering a particular KCKS model, this barotropic relation can be obtained (see the following subsection). 

Every solution $\{Q_0, \varphi(\tau)\}$ admits several thermodynamic interpretations. Each of these interpretations is settled by a different election of the two arbitrary real functions $s(Q), n(Q)$ in the thermodynamic schemes defined in (\ref{s-n-Tmodels})--(\ref{ell-m}). Now, these thermodynamic quantities depend on four parameters, $\{s_0, m_0, \ell_0, m_0 \}$, where $s_0 \equiv s(Q_0)$, $n_0 \equiv [N(Q_0)]^{-1}$, $\ell_0 \equiv \ell(Q_0)$, and $m_0 \equiv m(Q_0)$. The constant entropy of the isotropic evolution is $s_0$, and the matter density $n$ and the temperature are given by
\begin{equation}  \label{n-T-KCKS}
n = \frac{n_0 \varphi^{-3/2}}{\alpha_1 + \alpha_2 Q_0}\, , \qquad \Theta = \ell_0 \lambda_1(\tau) + m_0 \lambda_2(\tau) \, , \quad \lambda_i(\tau) \equiv \sqrt{\varphi}\left[\dot{\varphi} \dot{\alpha}_i + (\frac{\dot{\varphi}^2}{\varphi}- \ddot{\varphi}) \alpha_i\right] \, .
\end{equation}

\subsection{A Solution That Models an Ideal Gas in Isentropic Evolution} \label{sec-KCKS-ideal}

	In order to exemplify this discussion, let us consider the subfamily of the T-models compatible with the equation of state of a generic ideal gas $p = \tilde{k} n \Theta$. These \textit{ideal T-models} were studied in~\cite{FM-Termo-T-models} by imposing the ideal sonic condition (\ref{gas-ideal-sonic}). For the isentropic evolution we obtain a KCKS model of the form (\ref{metric-KCKS}) with $k=0$, $\alpha_1(\tau)=1$, and
\begin{equation} \label{phi-KCKS-ideal}
\varphi(\tau) = |\tau|^{\frac{4}{3 \tilde{\gamma}}}   ,  \qquad \alpha_2(\tau) =   |\tau|^{1- \frac{2}{\tilde{\gamma}}}  , \, \qquad  \quad  1< \tilde{\gamma} < 2 \, .
\end{equation}
The time coordinate takes values either in the interval $\tau>0$ (expanding models) or in the interval $\tau<0$ (contracting models). On the other hand, the expansion of the fluid flow takes the following expression:
\begin{equation}  \label{expansion-KCKS-ideal}
 \theta =\frac{2}{\tilde{\gamma}\,  \tau}  \Big(1 + \frac12 \delta \Big) ,
 \qquad \ \delta \! = \! \delta(\tau)\! \equiv \! \frac{(\tilde{\gamma}-2) \, Q_0 }{|\tau|^{\frac{2}{\tilde{\gamma}}-1} + Q_0}   \, ,
\end{equation}
and the pressure $p$ and the energy density $\rho$ take the following expression:
\begin{equation} \label{pressure-density-KCKS-ideal}
p = \frac{4(\tilde{\gamma} - 1)}{3 \tilde{\gamma}^2}\, \frac{1}{\tau^2} \, , \qquad \qquad
\rho = \frac{4}{3 \tilde{\gamma}^2}\, \frac{1}{\tau^2} [1 + \delta(\tau)] \, ,
\end{equation}
where $\delta(\tau)$ is given in (\ref{expansion-KCKS-ideal}). Note that we have a positive pressure. The case $Q_0=0$ leads to $\delta(\tau)=0$, and then we obtain a flat FLRW $\tilde{\gamma}$-law model: $p= (\tilde{\gamma} - 1) \rho$.

On the other hand, if we take $\tilde{\gamma}=1$, we obtain a dust solution, which is a non-isotropic generalization of the flat dust FLRW model. In fact, it is a Bianchi type I model that has been already considered~\cite{Hajj, Mazumder}. 

Nevertheless, our solutions for $\tilde{\gamma} \in ]1,2[$ are new Bianchi type I solutions that represent an ideal gas in isentropic evolution. In~\cite{FM-Termo-T-models}, we showed that these solutions fulfill the conditions for physical reality in wide spacetime domains. That study for the non-homogeneous case also apply here. Now, we analyze in more detail the physical behavior of the models. We only make explicit the expanding case ($\tau >0$). The analysis of the case $\tau <0$ is similar.

\begin{figure}[t]
				
				\parbox[c]{0.5\textwidth}{\includegraphics[width=0.47\textwidth]{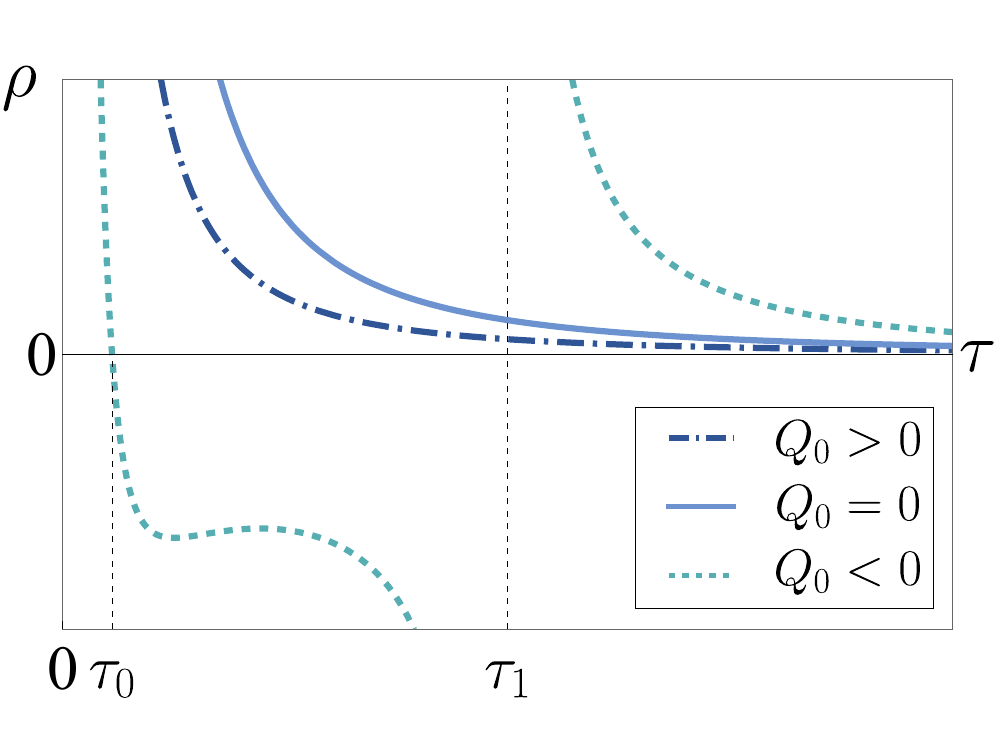}}
				\hspace{2pt}
				\parbox[c]{0.5\textwidth}{\includegraphics[width=0.47\textwidth]{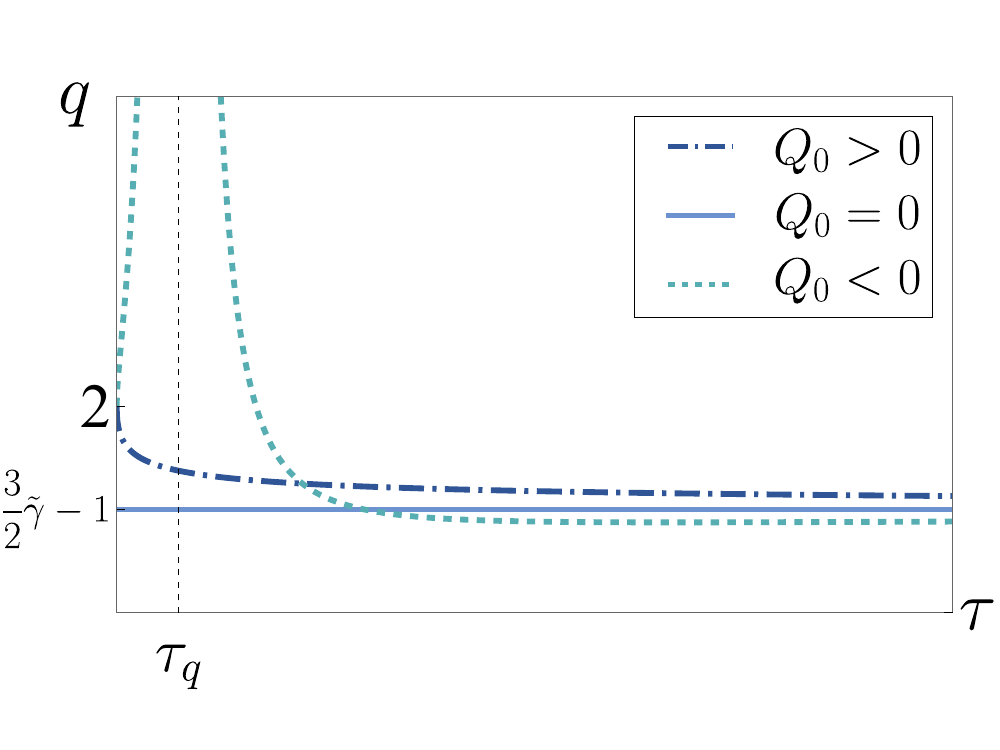}}
				\vspace{-3mm}
				\caption{At the \textbf{left}, evolution of the energy density $\rho$ of the ideal gas models for different values of $Q_0$. If $Q_0 > 0$, it is decreasing and positive everywhere, and if $Q_0 < 0$, it is decreasing and positive for $\tau > \tau_1$.  At the \textbf{right}, evolution of the average deceleration parameter $q$; it is positive and tends to $\frac32 \tilde{\gamma}-1$ for non-vanishing values of $Q_0$.}
				\label{Fig-1}
			\end{figure}

As we can see in Figure \ref{Fig-1}(left), the energy density is infinite at the big bang singularity $\tau = 0$ and decreases faster for non-vanishing values of $Q_0$ than for the FLRW limit $Q_0 = 0$. For negative values of $Q_0$ it becomes negative at $\tau = \tau_0 \equiv [-Q_0(\tilde{\gamma} - 1)]^\frac{\tilde{\gamma}}{2 - \tilde{\gamma}}$ and has another singularity at $\tau = \tau_1 \equiv (-Q_0)^\frac{\tilde{\gamma}}{2 - \tilde{\gamma}} > \tau_0$.

The average deceleration parameter $q \equiv -R \ddot{R} / \dot{R}^2$, where $R(\tau) \equiv [\varphi(\tau)]^{1/2} [1 + \alpha_2(\tau)\, Q_0]^{1/3}$ is the average scale factor, is positive everywhere (see Figure \ref{Fig-1}(right)). It is constant, $q =  \frac32 \tilde{\gamma}-1$, for the FLRW limit $Q_0 = 0$, and tends to this value for non-vanishing values of $Q_0$. Moreover, it has a singularity at $\tau = \tau_q \equiv (\frac{\tilde{\gamma}}{-2 Q_0})^\frac{\tilde{\gamma}}{2 - \tilde{\gamma}} < \tau_1$ if $Q_0 < 0$. 

\begin{figure}[H]

				\parbox[c]{0.5\textwidth}{\includegraphics[width=0.47\textwidth]{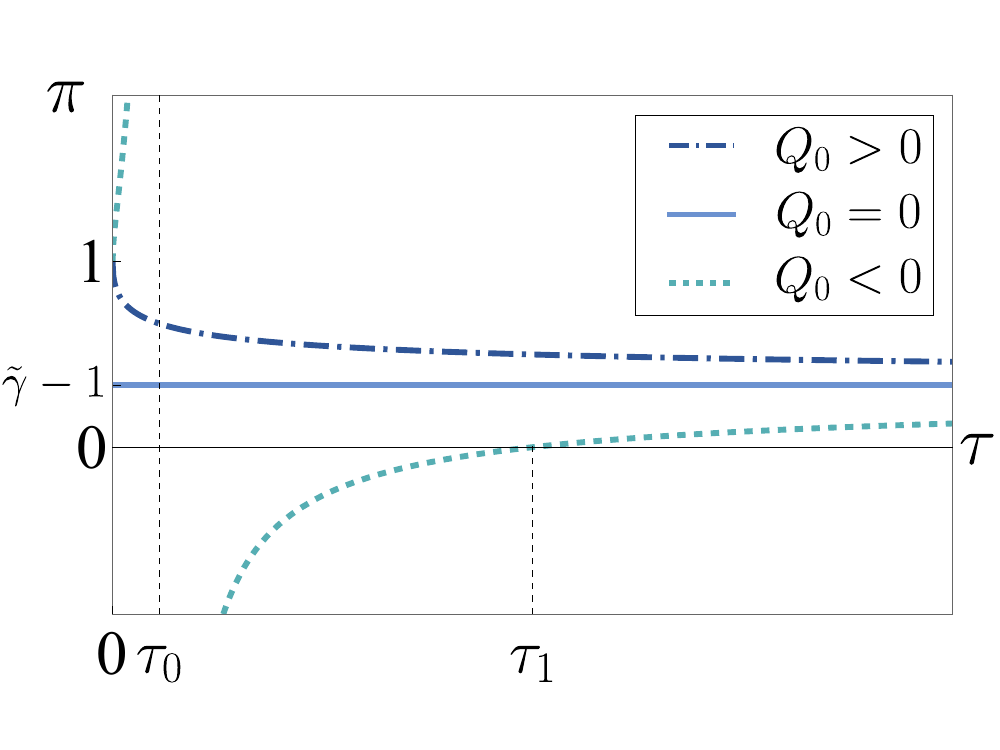}}
				\hspace{2pt}
				\parbox[c]{0.5\textwidth}{\includegraphics[width=0.47\textwidth]{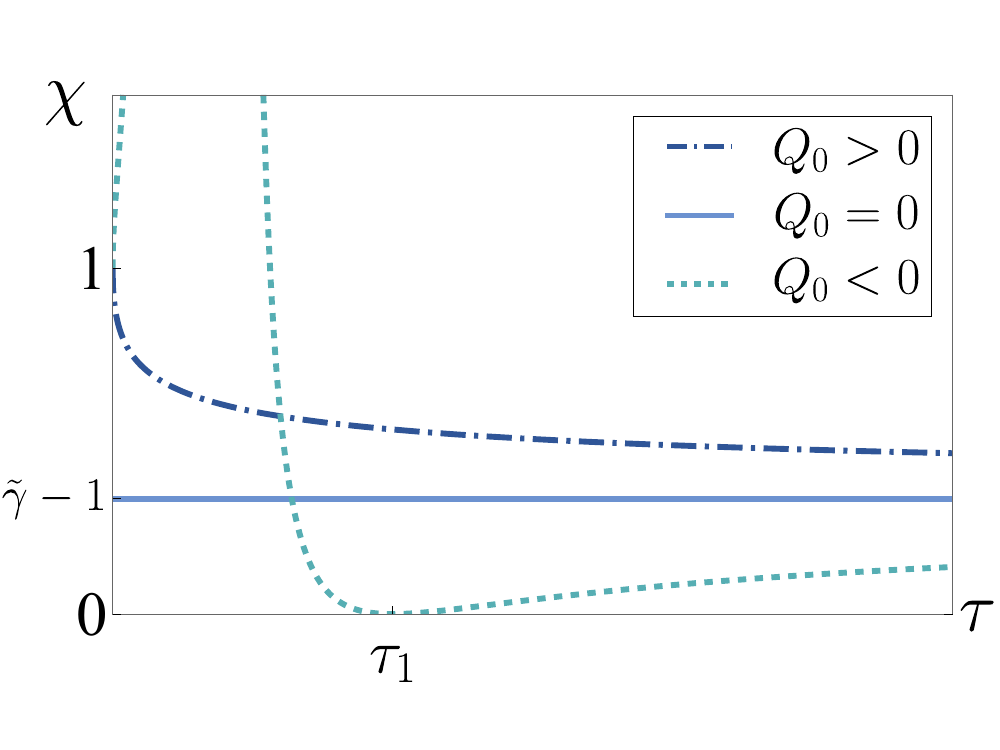}}
				\vspace{-3mm}
				\caption{At the \textbf{left}, evolution of the variable $\pi = p/\rho$ of the ideal gas models for different values of $Q_0$. At the \textbf{right}, evolution of the square of the speed of sound $\chi(\pi)$. Both hydrodynamic quantities have a similar behavior. If $Q_0 > 0$, they decrease in the interval $]1,\tilde{\gamma}-1[$ , and if $Q_0 < 0$, they increase in the interval $]0,\tilde{\gamma}-1[$ for $\tau > \tau_1$.}
				\label{Fig-2}
			\end{figure}

The hydrodynamic quantities $\pi = p/\rho$ and $\chi = u(p)/u(\rho)$ can be computed from (\ref{pressure-density-KCKS-ideal}), and we obtain
\begin{equation} \label{indicatrix-KCKS-ideal}
\pi = \frac{(\tilde{\gamma} - 1)(Q_0 + \tau^{\frac{2}{\tilde{\gamma}}-1})}{(\tilde{\gamma} - 1) \,Q_0 + \tau^{\frac{2}{\tilde{\gamma}}-1}}  \, , \qquad \qquad
\chi(\pi) = \frac{2 \tilde{\gamma} \pi^2}{(\pi+1)(\pi + \tilde{\gamma} -1)} \, .
\end{equation}
As we can see in Figure \ref{Fig-2}(left), the energy conditions hold in  the whole spacetime domain if $Q_0>0$. And if $Q_0 < 0$, the energy conditions are only fulfilled for $\tau > \tau_1$. Therefore, we will focus on this spacetime domain hereinafter for negative values of the parameter $Q_0$. Furthermore, the hydrodynamic compressibility conditions H$_1$ are also fulfilled in the region where the energy conditions hold (see Figure \ref{Fig-2}(right)).

\vspace{-9pt}

\begin{figure}[H]
				
				\parbox[c]{0.5\textwidth}{\includegraphics[width=0.47\textwidth]{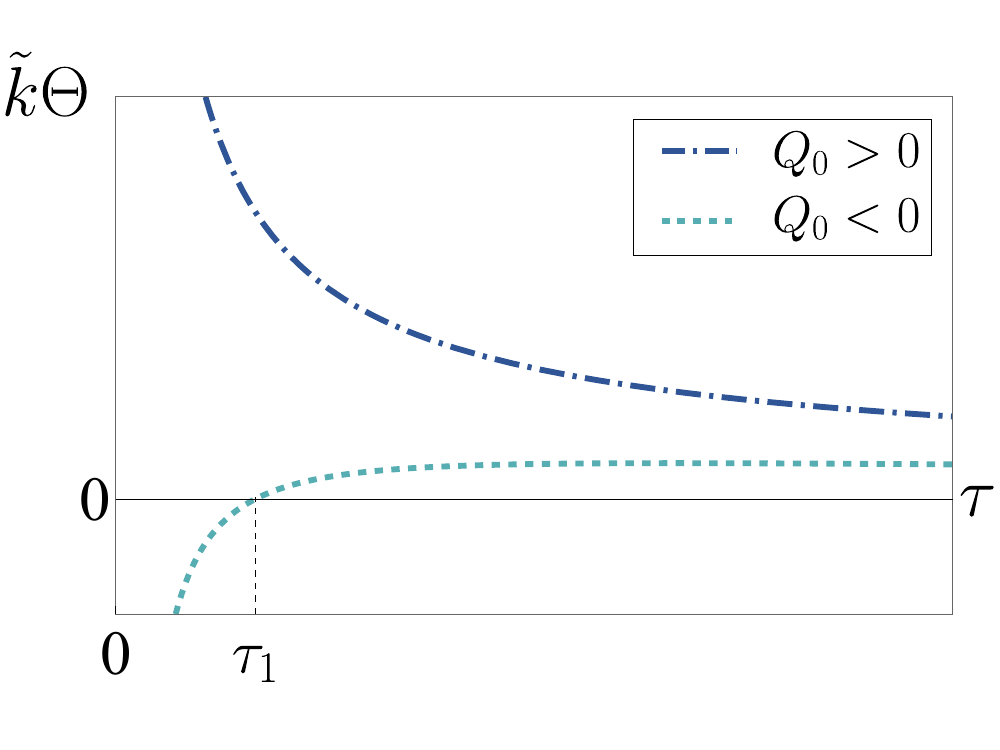}}
				\hspace{2pt}
				\parbox[c]{0.5\textwidth}{\includegraphics[width=0.47\textwidth]{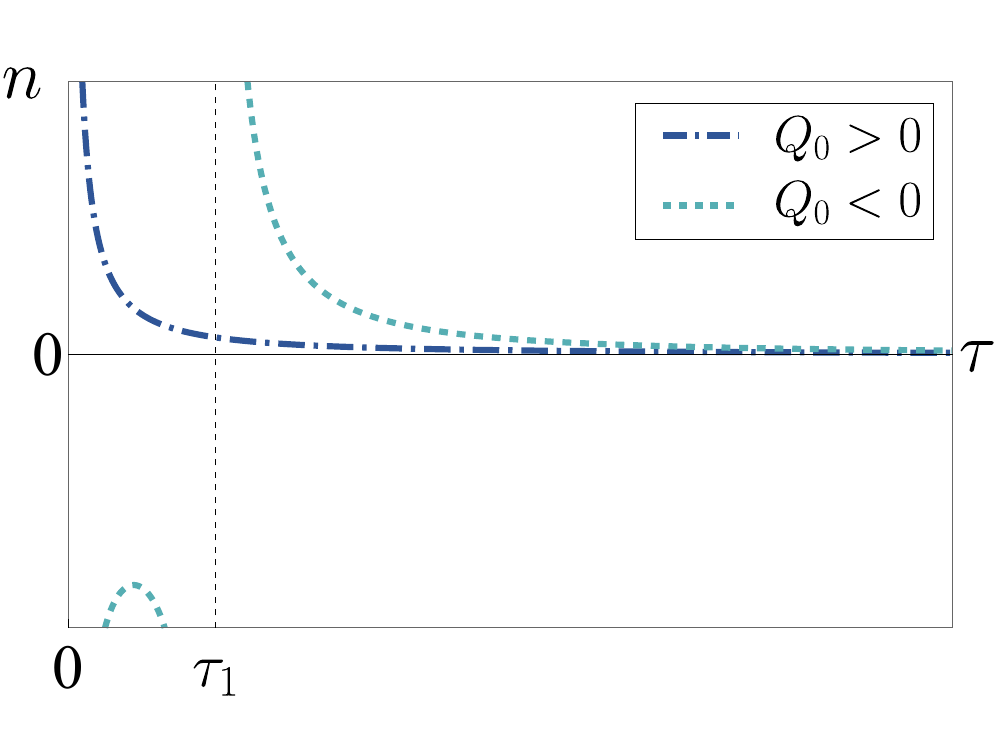}}
				\vspace{-3mm}
				\caption{At the \textbf{left}, evolution of the temperature $\Theta$ of the ideal gas models for the different values of $Q_0$. At the \textbf{right}, evolution of the matter density $n$.}
				\label{Fig-3}
			\end{figure}

The compatible thermodynamic schemes (\ref{n-T-KCKS}) have a constant entropy $s_0$, a matter density $n$, and a temperature $\Theta$, given by
\begin{equation}  \label{n-T-ideal-KCKS_a}
n = \frac{n_0 |\tau|^{-2/\tilde{\gamma}}}{1 + Q_0 \, |\tau|^{1-2/\tilde{\gamma}}}\, , \qquad \quad \Theta = \ell_0 \, \frac{4}{3\tilde{\gamma}} |\tau|^{-2+ 2/\tilde{\gamma}} + m_0 \, \frac{8(\tilde{\gamma}-1)}{3\tilde{\gamma}^2} \frac{1}{|\tau|}    \, .
\end{equation}
Which of the above thermodynamic schemes are compatible with the ideal gas equation of state? Note that, alternatively, we can determine these ideal schemes from the expressions (\ref{ideal-n-T}) and (\ref{ideal-s}), and we obtain the following (taking the constant $e_0$ such that $e(0)=1$):
\begin{equation} \label{n-T-ideal-KCKS_b}
n(\rho, p) = \rho (1\! -\! \pi)^{\frac{\tilde{\gamma}}{2-\tilde{\gamma}}} \,  \left[\frac{\tilde{\gamma}\!-\!1}{|\tilde{\gamma}\!-\!1\! -\! \pi|}\right]^{\frac{2(\tilde{\gamma}-1)}{2 - \tilde{\gamma}}}, 
\qquad \Theta(\rho, p)  = \frac{p}{\tilde{k} \, n(\rho, p)} \, ;
\end{equation}
\begin{equation} \label{s-ideal-KCKS}
s(\rho, p) = \bar{s}_0 + \tilde{k} \ln \left\{\frac1p \left|\frac{\tilde{\gamma}\!-\!1\! -\! \pi|}{1 - \pi}\right|^{\frac{2\tilde{\gamma}}{2 - \tilde{\gamma}}}\right\}, \quad \qquad \pi \equiv \frac{p}{\rho} \, .
\end{equation}
It is easy to prove that (\ref{n-T-ideal-KCKS_b}) are compatible with (\ref{n-T-ideal-KCKS_a}) if we consider in these expressions $n_0 = \frac{4 (2-\tilde{\gamma})}{3\tilde{\gamma}^2}Q_0^{-2\frac{\tilde{\gamma}-1}{2-\tilde{\gamma}}}$, $\ell_0 = \frac{\tilde{\gamma}-1}{\tilde{k}\tilde{\gamma}\, n_0}$, $m_0 = \frac{Q_0}{2 \tilde{k}\, n_0}$. Then, the temperature turns out to be
\begin{equation} \label{temperature-ideal-KCKS}
\Theta = q_0\, (|\tau|^{-2+ 2/\tilde{\gamma}} + \frac13 |\tau|^{-1}) , \qquad q_0 \equiv \frac{(\tilde{\gamma}-1)}{\tilde{k}(2-\tilde{\gamma})}Q_0^{2\frac{\tilde{\gamma}-1}{2-\tilde{\gamma}}}\, .
\end{equation}
For positive values of $Q_0$, the temperature is positive everywhere, being infinite at $\tau = 0$ and tending to zero at late times (see Figure \ref{Fig-3}(left)). For negative values of $Q_0$, it is positive for $\tau > \tau_1$, starting at zero, reaching a maximum and then tending to zero at late times, too. Regarding the matter density, it is positive everywhere if $Q_0 > 0$ and for $\tau > \tau_1$ if $Q_0 < 0$ (see Figure \ref{Fig-3}(right)). In both cases, it is infinite at the beginning and tends to zero at late times. It is worth remarking that the thermodynamic schemes (\ref{n-T-ideal-KCKS_a})--(\ref{s-ideal-KCKS}) do not apply in the FLRW limit $Q_0=0$. 

\begin{figure}[H]
				
				\hspace{-3pt}\parbox[c]{0.5\textwidth}{\includegraphics[width=0.46\textwidth]{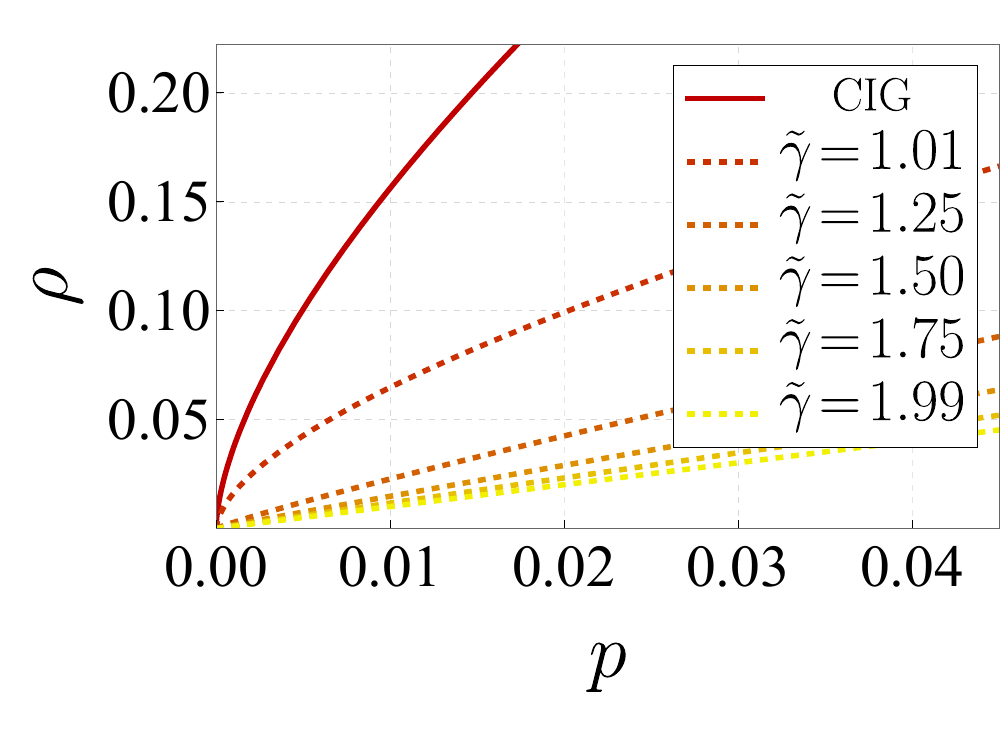}}
				\hspace{2pt}
				\parbox[c]{0.5\textwidth}{\includegraphics[width=0.46\textwidth]{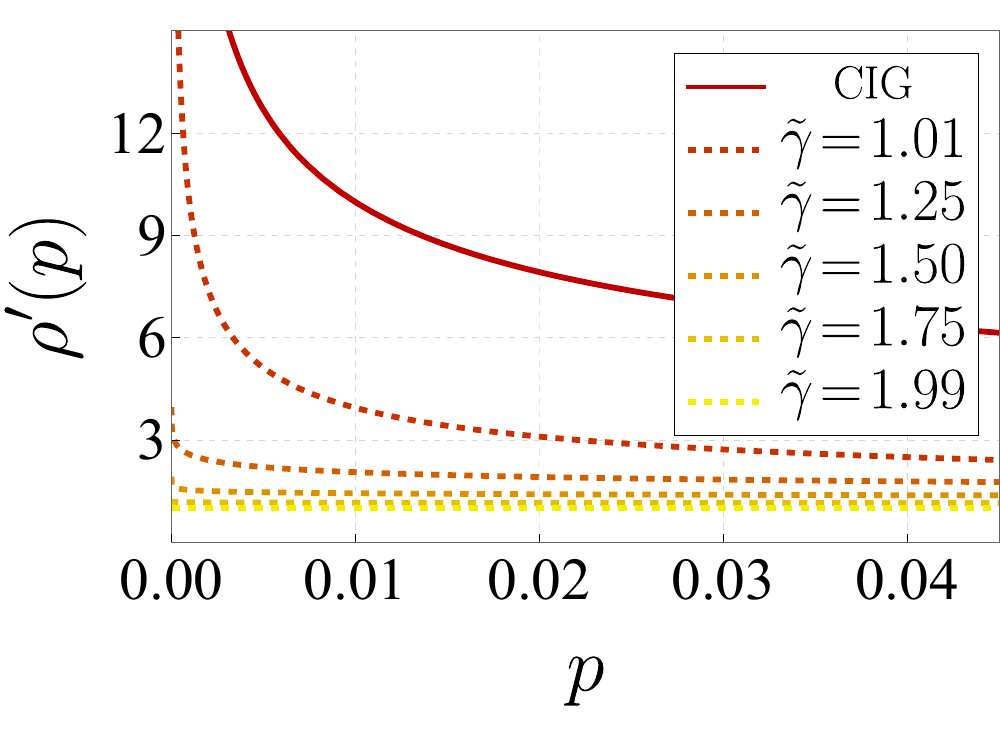}}
				\vspace{-4mm}
				\caption{Comparison of the barotropic relation $\rho = \rho (p)$ (\textbf{left}) and its first derivative (\textbf{right}), for a classical ideal gas with $\gamma = 5/3$ and for different KCKS ideal models.}
				\label{Fig-4}
			\end{figure}

Moreover, the isentropic evolution, $s(\rho, p) = s_0$, implies from (\ref{s-ideal-KCKS}) (or by using (\ref{expansion-KCKS-ideal}) and (\ref{pressure-density-KCKS-ideal})) that in this case we have the following barotropic relation:
\begin{equation} \label{rel-barotropia-T-model-ideal-isentropic}
\rho = p \frac{1 + \tilde{B} \, p^{\frac{2 - \tilde{\gamma}}{2\tilde{\gamma}}}}{(\tilde{\gamma}\! -\! 1) + \tilde{B} \, p^{\frac{2 - \tilde{\gamma}}{2\tilde{\gamma}}}} \equiv \rho (p)  , \quad \tilde{B} \equiv \exp\left\{\!\frac{2\! - \! \tilde{\gamma}}{2\tilde{\gamma}\tilde{k}}(s_0\! - \! \bar{s}_0)\!\right\} \!=\! Q_0 (\tilde{\gamma}\!-\!1)\left[\frac{3  \tilde{\gamma}^2}{4(\tilde{\gamma}\!-\!1)}\right]^{\frac{2 \tilde{\gamma}}{2\!-\!\tilde{\gamma}}}\! \!.
\end{equation}
This barotropic relation behaves, at low temperatures, as that of a classical ideal gas (\ref{CIG-isentropic}) at zero order (see Figure \ref{Fig-4}(left)). At first order, it approaches better than that of a classical ideal gas with $\gamma = 5/3$ the closer $\tilde{\gamma}$ is to $1$, but never reaches its behavior (see Figure \ref{Fig-4}(right)).
			%

	
		\section{Other Interpretations} \label{sec-KCKS-other}
		
		In the previous section, we saw that every KCKS solution performs the isentropic evolution of the fluids represented by a T-model. However, it might also be the case that these solutions also represent the isentropic evolution of other families of fluids. 
		
Two simple fluids with different equation of state, $p = p_1(\rho, s)$ and $p = p_2(\rho, s)$, can have an isentropic evolution defining the same barotropic relation $p = \psi(\rho) = p_1(\rho, s_0)  = p_2(\rho, s_0)$. Therefore, by imposing a specific barotropic relation, we can obtain the KCKS subfamilies representing the isentropic evolution of more general non-barotropic families of fluids fulfilling such barotropic relations along the evolution.

			\subsection{Isentropic Evolution of a Classical Ideal Gas}
			
			A CIG in isentropic evolution fulfills the barotropic relation (\ref{CIG-isentropic}). Integrating the conservation of matter Equation (\ref{c-masa}) with $n$ given in (\ref{n-CIG}) and $\theta$ in (\ref{expansion-T-1}), we get $n = n_0 \left(\frac{R_0}{R}\right)^3$, with $n_0$ constant and $R \equiv (\varphi^{3/2} \alpha)^{1/3}$ the average scale factor. With that, we can write the pressure as a function of the metric functions for this case as
			\begin{equation} \label{p-GIC-isentropic}
				p = p_0 (\varphi^{3/2} \, \alpha)^{-\gamma}, \qquad p_0 = 
				constant \, ,
			\end{equation}
where $\gamma$ is the adiabatic index of the CIG. The energy density $\rho$, is given by (\ref{density-T-1}) with $\alpha = \alpha_1 (t) + \alpha_2 (t)  Q_0$, and $Q_0$ constant. Thus, plugging this expression and (\ref{p-GIC-isentropic}) into the barotropic relation (\ref{CIG-isentropic}), we get the following differential equation for the metric functions of a KCKS solution representing the isentropic evolution of a non-barotropic family of solutions fulfilling the CIG barotropic relation along the evolution:
			\begin{equation} \label{rel-barotropia-KCKS-GIC-isentropic}
				\bar{n}_0 [\varphi^{3/2} (\alpha_1\! +\! \alpha_2 \, Q_0)]^{-1}\! +\! \frac{\bar{p}_0}{\gamma - 1} [\varphi^{3/2} (\alpha_1 \! +\! \alpha_2 \, Q_0)]^{-\gamma}\! = v \left( \frac34 \frac{\dot{\varphi}^2}{\varphi^2}\! +\! \frac{\dot{\varphi}}{\varphi} \frac{\dot{\alpha}_1\! +\! \dot{\alpha}_2 \, Q_0}{\alpha_1\! +\! \alpha_2 \, Q_0} \right)\! +\! \frac{k}{\varphi} \, ,
			\end{equation}
where $\bar{n}_0$ and $\bar{p}_0$ are positive constants.

			It is worth remarking that the metric functions $\{\varphi, v, \alpha_i\}$ still need to fulfill the perfect fluid field equations (\ref{eq-T-1}). Thus, Equation (\ref{rel-barotropia-KCKS-GIC-isentropic}), together with the field equations obtained by imposing (\ref{eq-T-1}) on each of the functions $\alpha_i$, constitute a set of three constraints on four metric functions. We can now use the freedom to choose the time coordinate to impose a fourth equation that closes the system. In~\cite{FM-General-sol-T-models}, we already used three of these four restrictions to write three of the metric functions in terms of the fourth one. Thus, we can now use those results to write (\ref{rel-barotropia-KCKS-GIC-isentropic}) as a differential equation for the only remaining metric~\mbox{function}. 
			
			In particular, for the case $k = 0$, the modified Herlt algorithm~\cite{FM-General-sol-T-models} uses the field equations to give $v = \varphi^{-3}$ and $\alpha = 1 + t \, Q_0$. Then, the metric tensor becomes
\begin{equation} \label{metric-KCKS-CIG}
ds^2= - \varphi^{-3}(t)\, d t^2 + \varphi(t)[(1 + t \,Q_0)^2 dr^2 +  dx^2+dy^2].
\end{equation}
Moreover, by substituting this into (\ref{rel-barotropia-KCKS-GIC-isentropic}), we get a single first-order differential equation for the metric function $\varphi(t)$:
			\begin{equation} \label{rel-barotropia-KCKS-GIC-isentropic-k_0}
				\bar{n}_0 [\varphi^{3/2} (1 + t \, Q_0)]^{-1} + \frac{\bar{p}_0}{\gamma - 1} [\varphi^{3/2} (1 + t \, Q_0)]^{-\gamma} = \frac{1}{\varphi^3} \left( \frac34 \frac{\dot{\varphi}^2}{\varphi^2} + \frac{\dot{\varphi}}{\varphi} \frac{Q_0}{1 + t \, Q_0} \right) \, .
			\end{equation}

\begin{figure}[t]
				\parbox[c]{0.50\textwidth}{\includegraphics[width=0.5\textwidth]{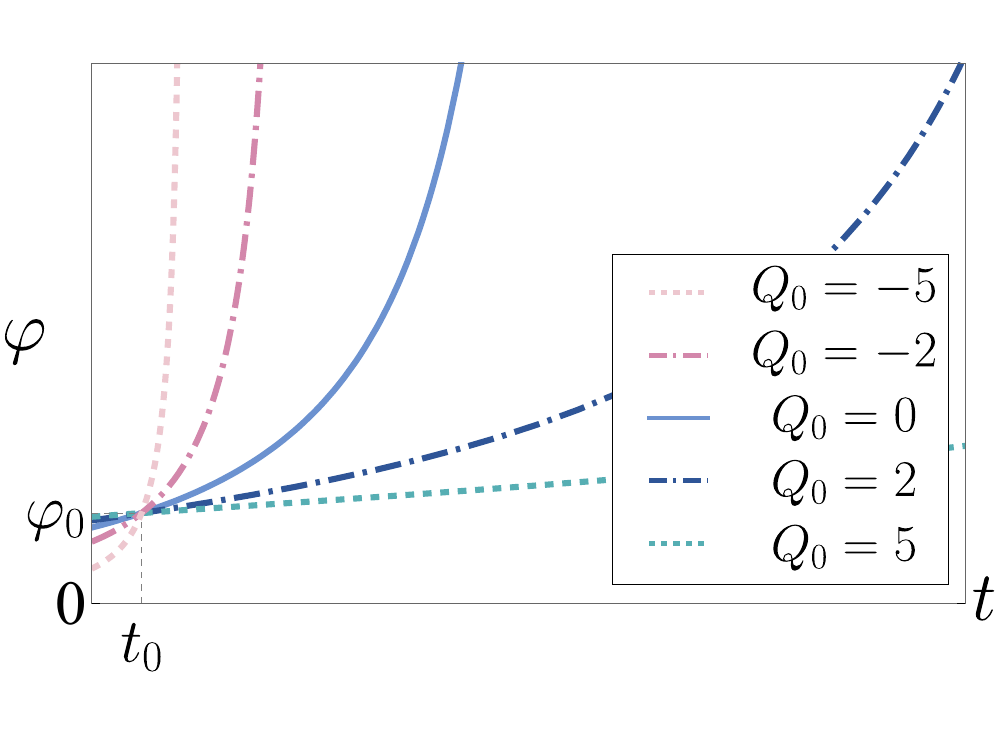}}
				\hspace{2pt}
				\parbox[c]{0.50\textwidth}{\includegraphics[width=0.5\textwidth]{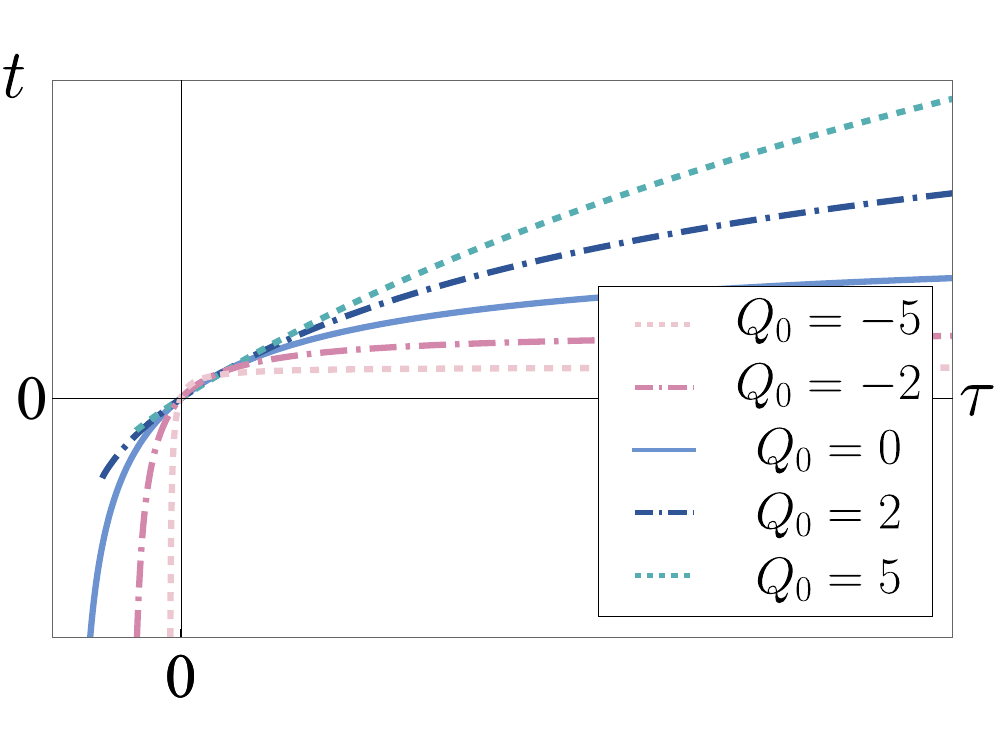}}
				\vspace{-4mm}
				\caption{At the \textbf{left}, comparison of the evolution of the solution $\varphi(t)$ of the differential Equation (\ref{rel-barotropia-KCKS-GIC-isentropic-k_0}) with $\gamma = 5/3$ (monoatomic case) and initial condition $\varphi(t_0) = \varphi_0$ for different values of the constant parameter $Q_0$. At the \textbf{right}, the same comparison for the function relating the time coordinate $t = \alpha_2$ with the proper time $\tau$.}
				\label{Fig-5}
			\end{figure}

This differential equation needs to be solved numerically for different values of the constant parameter $Q_0$ and the adiabatic index $\gamma$. In the left plot of Figure \ref{Fig-5} we show the solution $\varphi(t)$ for the monoatomic case $\gamma = 5/3$ and different values of $Q_0$. It is worth remarking that we recover the FLRW solutions in the limit $Q_0 = 0$. This isotropic case was already considered in~\cite{CFS-CIG} by taking the co-moving proper time. Recall, however, that, here, we are taking $\alpha_2$ as the time coordinate. 

If we want to compare these solutions with the well-known results for the FLRW metrics, we need to plot these quantities as functions of the proper time $\tau = \int \varphi(t)^{3/2} \textrm{d}t = \tau(t)$. This can be achieved after performing this integral and inverting the resulting function numerically, which gives the time coordinate as a function of the proper time, $t = t(\tau)$ (see the right plot of Figure \ref{Fig-5} for the monoatomic case $\gamma = 5/3$ and for different values of $Q_0$). It is important to note that the function $t(\tau)$ tends to a finite value $ t_M$ when $\tau$ tends to infinity, with $t_M$ being bigger for greater values of $Q_0$, and that $t(0) = 0$. 

			\begin{figure}[H]
			\parbox[c]{0.5\textwidth}{\includegraphics[width=0.5\textwidth]{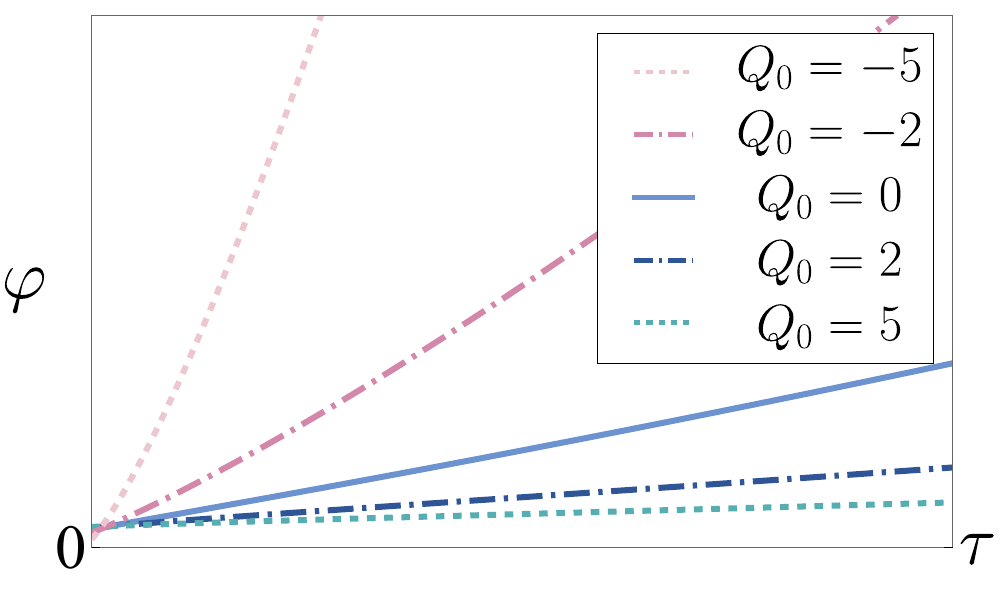}}
			\hspace{8pt}
			\parbox[c]{0.5\textwidth}{\includegraphics[width=0.5\textwidth]{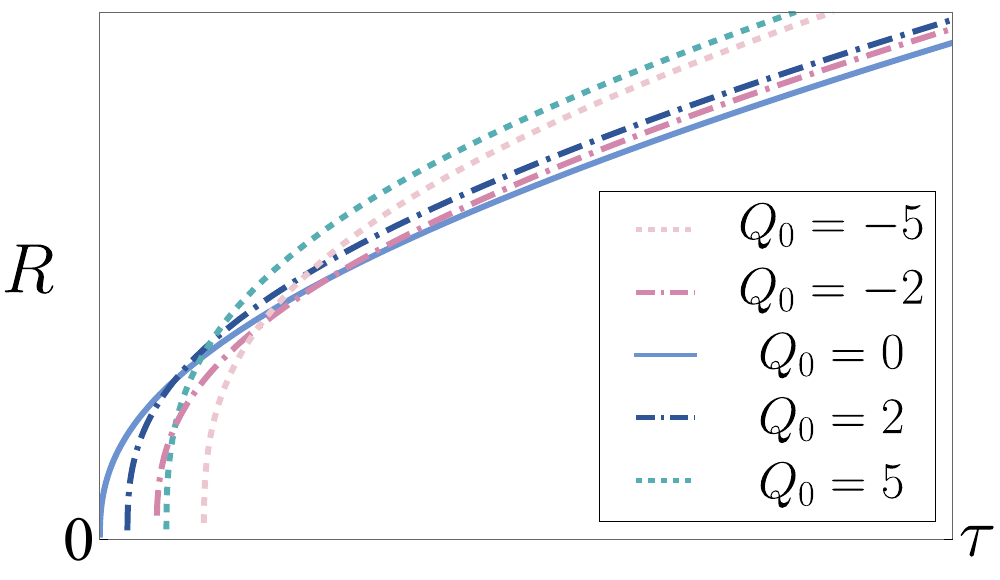}}
			\caption{At the \textbf{left}, comparison of the evolution of the metric function $\varphi(\tau)$ with $\gamma = 5/3$ (monoatomic case) for different values of the constant parameter $Q_0$. At the \textbf{right}, the same comparison for the average scale factor $R(\tau)$.}
			\label{Fig-6}
			\end{figure}

The left plot of Figure \ref{Fig-6} shows the solution $\varphi(\tau)$ of the differential Equation (\ref{rel-barotropia-KCKS-GIC-isentropic-k_0}) in terms of the proper time,  for the monoatomic case $\gamma = 5/3$ and for different values of $Q_0$. Notice that $\varphi(\tau)$ grows faster the smaller $Q_0$ is. The right plot of Figure \ref{Fig-6} shows the average scale factor $R(\tau) \equiv [\varphi(\tau)]^{1/2} [1 + t(\tau)\, Q_0]^{1/3}$ that results for different values of $Q_0$. We can see that the effect of the existence of the constant $Q_0 \neq 0$ is to make the function $R$ grow faster compared with the FLRW case $Q_0 = 0$. Moreover, for a given sign of $Q_0$, the bigger the value of $|Q_0|$, the faster $R(\tau)$ grows, while, for late times, the behavior of the function is determined by this absolute value. Regarding the average deceleration parameter, it is positive and decreasing everywhere. Furthermore, its behavior is soon dominated by the value of $|Q_0|$, being greater for bigger values of $|Q_0|$.

\begin{figure}[H]
			\parbox[c]{0.5\textwidth}{\includegraphics[width=0.5\textwidth]{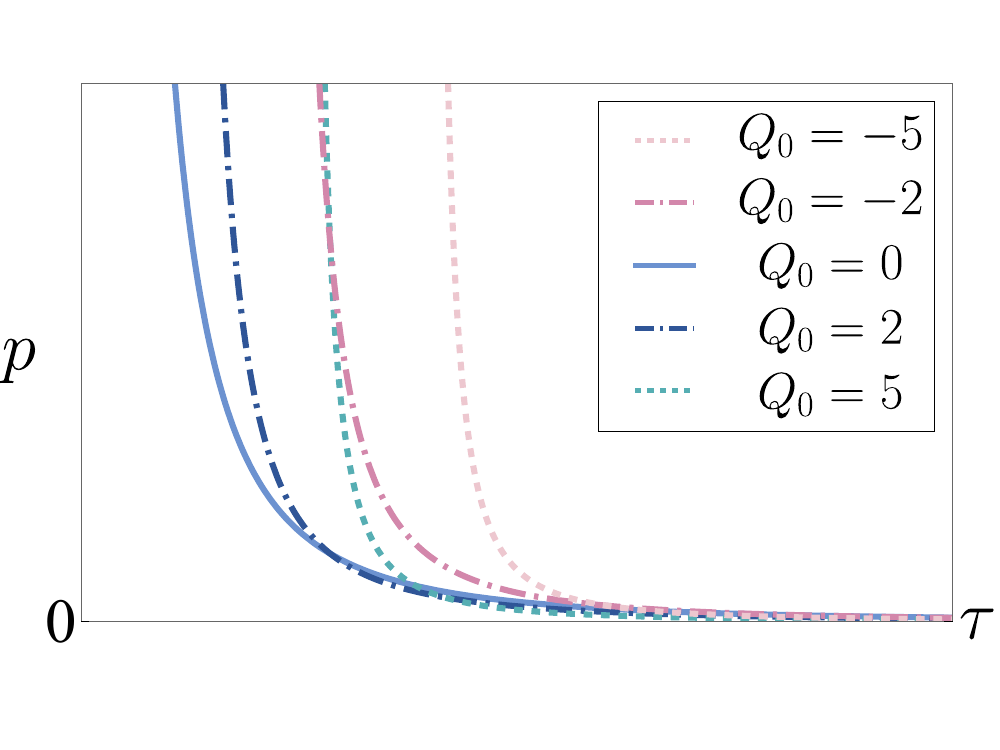}}
			\hspace{8pt}
			\parbox[c]{0.5\textwidth}{\includegraphics[width=0.5\textwidth]{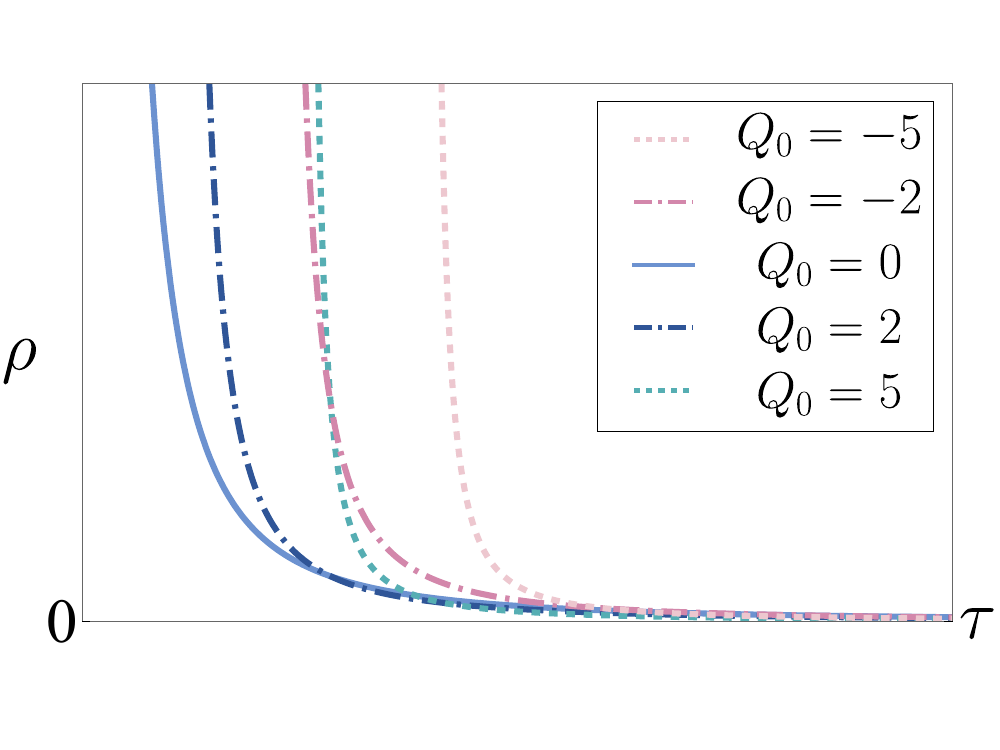}}
			\vspace{-6mm}
			
			\caption{At the \textbf{left}, comparison of the evolution of the pressure $p(\tau)$ with $\gamma = 5/3$ (monoatomic case) for different values of the constant parameter $Q_0$. At the \textbf{right}, the same comparison for the energy density $\rho(\tau)$. }
			\label{Fig-7}
			\end{figure}

It is worth remarking that, for $\gamma \in [1,2[$, the whole energy and compressibility conditions hold on the spacetime regions where the pressure is positive~\cite{CFS-CIG}. In the left plot of Figure \ref{Fig-7}, we show the behavior of the pressure. Note that it is a positive decreasing function (which decreases faster the bigger the value of $|Q_0|$ is) that vanishes at infinity. Therefore, all the energy and compressibility conditions hold on the whole spacetime region where the solution is defined. The behavior of the energy density is shown in the right plot of Figure \ref{Fig-7} and is qualitatively similar to that of the pressure.

\vspace{-12pt}

As mentioned above, the FLRW limit $Q_0 = 0$ has already been  considered~\cite{CFS-CIG}. The solutions with $Q_0 \not=0$ are new Bianchi type I models that represent a classical ideal gas in isentropic evolution.

			\subsection{$\gamma$-Law Barotropic Relation}

Another barotropic relation usually considered in the literature is the so-called $\gamma$-law, $p = (\gamma\! -\! 1)\rho$. Its thermodynamic interpretation is unclear except for the case $\gamma =4/3$, which models an ultrarelativistic fluid. The FLRW models with a $\gamma$-law were studied in~\cite{Assad-Lima}, and several authors have partially integrated Einstein equations for the KCKS metrics~\cite{Harrison, Vajk} (see also~\cite{Kramer}). 

Now, we present the general solution for an arbitrary $\gamma$ in the case $k=0$. Again, we use the algorithm introduced in~\cite{FM-General-sol-T-models}. We take $v = \varphi^{-3}$ and $\alpha = 1 + t \, Q_0$. Then, the metric tensor takes the expression (\ref{metric-KCKS-CIG}), and the $\gamma$-law condition becomes a differential equation for the metric function $\varphi(t)$:
\begin{equation}
\frac{\ddot{\varphi}}{\dot{\varphi}} + \frac14 (3\gamma - 10) \frac{\dot{\varphi}}{\varphi} + (\gamma - 1) \frac{Q_0}{1 + Q_0 t} = 0 \, .
\end{equation}
This equation can be integrated, and for $Q_0 \not=0$ the general solution is
\begin{equation}
\varphi(t) = \left[D + (1+Q_0 t)^{2-\gamma}\right]^\frac{4}{3(\gamma-2)} \, , \qquad D = constant .
\end{equation}
The study of the physical behavior of the solutions implies obtaining the thermodynamic quantities in terms of the proper time, which requires the numerical determination of $t(\tau)$.

Now we focus on the physically relevant case of a radiation fluid, that is, $\gamma = 4/3$. Then, it is easier to work with the proper time  ($v=1$), and the metric tensor takes the following~\mbox{expression}:
\begin{equation} \label{metric-KCKS-gamma}
ds^2= -d \tau^2 + Y^2(\tau)[\alpha^2(\tau)  dr^2 +  dx^2+dy^2] .
\end{equation}
Moreover, the $\gamma$-law condition and the field equation are, respectively,
\begin{equation} \label{eq-gamma}
\frac{\ddot{\varphi}}{\dot{\varphi}} + (\frac34 \gamma - 1) \frac{\dot{\varphi}}{\varphi} + (\gamma - 1) \frac{\dot{\alpha}}{\alpha} = 0 \, , \qquad 2 \varphi \ddot{\alpha} + 3 \dot{\varphi} \dot{\alpha} = 0 ,
\end{equation} 
where $\varphi = Y^2$. These equations can be partially integrated, and we obtain
\begin{equation}
\dot{\varphi} \varphi^{\frac34 \gamma - 1}\alpha^{\gamma-1} = \tilde{C}, \qquad  \dot{\alpha} \varphi^{3/2}= \tilde{K} , \qquad \tilde{C}, \tilde{K} = constants.
\end{equation} 
Using these expressions, we can remove $\alpha$ and $\dot{\alpha}$ from the first equation in (\ref{eq-gamma}), and we obtain a differential equation for the sole metric function $\varphi(\tau)$:
\begin{equation} \label{eq-gamma-b}
\frac{\ddot{\varphi}}{\dot{\varphi}} + (\frac34 \gamma - 1) \frac{\dot{\varphi}}{\varphi} + \hat{C} \dot{\varphi}^{\frac{1}{\gamma-1}} \varphi^{\frac{2-3 \gamma}{4(\gamma-1)}} = 0 \, \qquad \hat{C} =constant.
\end{equation} 
For $\gamma = 4/3$, this equation is equivalent to $\dot{\varphi}^{-2} + 4 \hat{C} \varphi^{-1/2} = \hat{K}$, $\hat{K}= constant$, which, in terms of $Y=\sqrt{\varphi}$, writes
\begin{equation} \label{Y_punt}
\dot{Y} = [K Y^2 + C Y]^{-\frac{1}{2}}, \qquad K, C = constants \, .
\end{equation} 
The solution to this equation determines a KCKS model representing a radiation fluid. Note that the change of the metric functions $Y(\tau)$ and $\alpha(\tau)$ by a constant factor does not change the tensor metric (\ref{metric-KCKS-gamma}) as we can change the coordinates $r,x,y$ by a factor. Moreover, the proper time $\tau$ can be changed by an additive constant. We take into account these facts in solving Equation (\ref{Y_punt}). 

If $C=0$, the solution is $Y(\tau) = \sqrt{\tau}$, $\alpha(\tau)=1$, and we obtain a dominant radiation flat FLRW model. If $C \not=0$ and $K=0$, the solution to (\ref{Y_punt}) leads to
\begin{equation}
Y(\tau) = \tau^{2/3} \,  , \qquad \alpha(\tau) = 1/\tau \, ,
\end{equation}
which is the particular case of the vacuum Kasner models corresponding to the non-static region of the A$_3$-metric.

Finally, if $C \not=0 \not= K$, Equation (\ref{Y_punt}) leads to the following solutions:
\begin{eqnarray}
\hspace{-4mm} K>0: \qquad   \tau(Y) = A[(Y\!+\!1) \sqrt{2Y \!+ \!Y^2} - {\rm arcch} (Y\!+\!1)],\; \quad \alpha(\tau)\! =\! \left(\frac{2}{Y(\tau)}\!+\!1\right)^{3/2}\!\!, \\
\hspace{-4mm} K<0: \qquad   \tau(Y) = B[(Y\!-\!1) \sqrt{2Y \! -\!Y^2} - \arccos (Y\!-\!1)], \quad \alpha(\tau)\! =\! \left(\frac{2}{Y(\tau)}\!-\!1\right)^{3/2}\!\!,
\end{eqnarray} 
where $A,B = constants$.


\section{Ending Comments}
\label{sec-conclusions}

The physical meaning of the KCKS solutions has always been questioned. However, in this work, we have shown that they can be interpreted as a fluid in local thermal equilibrium. More specifically, we have seen here that each KCKS metric models a thermodynamic fluid in isentropic evolution. Moreover, some of them can represent a generic ideal gas.

We have also analyzed the solutions that model a classical ideal gas, as well as the equations for the $\gamma$-law models. 

Although our general approach and results apply to models with any symmetry (spherical, plane, or hyperbolic), here we have studied in more detail the case of plane symmetry, which allows a more in-depth analysis of the solutions. These plane KCKS solutions are Bianchi type I cosmologies, adding new physically realistic models to already known solutions. Further research should be devoted to extending the considered solutions to the case of general Bianchi type I models (with three different scale factors).

%
%
%
%
%

\vspace{6pt} 




\authorcontributions{S.M. and J.J.F. contributed equally to develop the idea of the manuscript. Both authors have read and agreed to the published version of the manuscript.}

\funding{This work was supported by the Generalitat Valenciana Project CIAICO/2022/252 and the Plan Recuperaci\'on, Transformaci\'on y Resiliencia, project ASFAE/2022/001, with funding from European Union NextGenerationEU (PRTR-C17.I1).}

\institutionalreview{Not applicable. 
}

\informedconsent{Not applicable. 
}

\dataavailability{Not applicable. 
} 

\acknowledgments{S.M. acknowledges financial support from the Generalitat Valenciana (grant CIACIF/2021/028). We would like to thank the /Universe/ Editorial Office at MDPI for the invitation to submit this manuscript free of charge.}

\conflictsofinterest{The authors declare no conflicts of interest.} 






\begin{adjustwidth}{-\extralength}{0cm}

\reftitle{References}

\PublishersNote{}
\end{adjustwidth}

\end{document}